\documentclass[a4paper,12pt,english,german]{article}

\usepackage{mathtools}
\usepackage{graphicx}

\usepackage{amssymb}

\usepackage{amsmath}

\begin{document}

\centerline{\bf Dynamical emergence of Markovianity in Local Time
Scheme}

\centerline{
J. Jekni\' c-Dugi\' c$^{1}$, M. Arsenijevi\' c$^{2}$ and M. Dugi\'
c$^{2}$}

$^{1}$Department of Physics, Faculty of Sciences and Mathematics,
18000 Ni\v s, Serbia

 $^{2}$Department of Physics, Faculty of
Science, 34000 Kragujevac, Serbia

\centerline{Markovian processes; Quantum dynamical maps; Local
time}

\bigskip

{\bf Abstract} Recently we pointed out the so-called Local Time
Scheme  as a novel approach to quantum foundations that solves the
preferred pointer-basis problem. In this paper we introduce and
analyze in depth a rather non-standard dynamical map that is
imposed by the scheme. On one hand, the map does not allow for
introducing a properly defined generator of the evolution nor does
it represent a quantum channel. On the other hand, the map is
linear, positive, trace preserving and unital as well as
completely positive, but is not divisible and therefore
non-Markovian. Nevertheless, we provide quantitative criteria for
dynamical emergence of time-coarse-grained Markovianity, for exact
dynamics of an open system, as well as for operationally-defined
approximation of a closed or open many-particle system.  A closed
system never reaches a steady state, while an open system may
reach a unique steady state given by the L\" uders-von Neumann
formula; where the smaller the open system, the faster a steady
state is attained. These generic findings extend the standard open
quantum systems theory and substantially tackle certain
cosmological issues.

\bigskip

{\bf 1. Introduction}

\bigskip

Recently we pointed out the so-called Local Time Scheme [1] as a
novel non-interpretational, minimalist approach to quantum
foundations. In Local Time Scheme, dynamics is a primitive that
asymptotically defines local time for a single closed ('local')
quantum system [1, 2]. In general, quantum systems are subjected
to different local times and hence there is not uniquely defined
time for an [statistical] ensemble of such systems as well as for
the Universe as a whole.

\noindent Instead of the universal ('global') time we learn about
a single closed system's local time as a hidden classical
parameter of the system's dynamics. This 'multi-time' scheme
establishes dynamical change of a single-system's local time when
the system sufficiently-strongly interacts with another system
that is not subjected to the same local time.
 The scheme [1] naturally differentiates between the few- and
many- particle systems, routinely and technically-simply describes
quantum measurement, resolves the 'preferred pointer-basis
problem' for an ensemble of bipartitions of closed many-particle
systems and provides a plausible interpretation of the
Wheeler-DeWitt equation.

Understanding Time is a deep issue of physics and philosophy.
Nevertheless, resolving this issue is not yet necessary in the
Local Time Scheme. Moreover, mathematical formalization [1] of the
concept of local time is an unexpected tool for distinguishing
between the few- and many-particle systems and hence may be
expected  to concern the whole set of the foundational,
interpretation-related issues in quantum theory such as quantum
measurement, microscopic origin of the phenomenological 'arrow of
time' and the problem of the 'transition from quantum to
classical', which, in turn, is recognized as the ultimate basis of
the new technologies [3-5].

In contrast to LTS,  quantum foundational research is often
fragmented and typically goes to interpretations [6-8] or
completely discards some foundational problems in a purely
operational manner [9]. Crosstalk is exceptional and has only
recently started in more systematic and comprehensive forms
[10-14]. The outcome of this new endeavor is  hard to predict.

In this paper we do not tackle the interpretation-related issues
but rather pursue the so-far-useful methodology of Ref. [1] to
address the following question: Whether or not the scheme
introduces some new elements or insights into the standard theory
of open quantum systems [15-18]? As a remote goal we recognize
value addressing the following question: What might be the
consequences of non-unique time in regard of certain basic issues
in cosmology?

As a step forward in investigating the implications of the idea of
local time, in the spirit of [1], our considerations are
minimalist and resort to the mathematical aspects that constitute
the basis for the future investigation of the interpretational
implications of Local Time Scheme. Bearing in mind importance of
Markovian processes [15-17], we investigate (non)Markovian
character of the dynamics imposed by LTS. To this end, we stick to
the following definition that is essentially taken over from Ref.
[17]:

\noindent {\bf Def.1.1} {\it A quantum system is said to undergo
Markovian dynamics if its dynamics is described by a family of
dynamical maps}, $\{\mathcal{E}_{(t_2, t_1)}, t_2 \ge t_1 \}$ {\it
such that, for every}  $t_2 \ge t_1$, $\mathcal{E}_{(t_2, t_1)}$
{\it is a completely positive map and fulfils the composition law}
$\mathcal{E}_{(t_3, t_1)} = \mathcal{E}_{(t_3,
t_2)}\mathcal{E}_{(t_2, t_1)}, t_3 \ge t_2 \ge t_1$.

The following are the main findings of this paper: (i) For a
closed many-particle system, the map is completely positive but
neither divisible nor differentiable; (ii) Regarding an observer
not capable of resolving the close energy values of a closed
many-particle system, the (approximate) map is divisible and
dynamically acquires complete positivity and hence
time-coarse-grained Markovianity (TCGM); (iii) Depending on the
system's energy, an observer can in principle detect low-energy
(e.g. low temperature) Markovian dynamics, the unitary-like
Markovian behavior for relatively high energies (e.g. high
temperature), and non-Markovian dynamics for the rest of quantum
states of a closed many-particle system; (iv) Regarding an open
system in a proper strong interaction with a many-particle
environment, the exact map is completely positive and dynamically
acquires divisibility and hence TCGM; (v) An approximate map for
the open system is divisible and dynamically acquires complete
positivity and therefore TCGM;  (vi) A closed system never reaches
a steady state, while an open system may reach unique steady
state, which is given by the L\" uders-von Neumann formula (i.e.
by the von Neumann's projection postulate) in quantum measurement;
(vii) The smaller the open system the faster is reached the steady
state.

In the context of the new fundamental dynamical law [that takes
the place and the role of the standard Schr\" odinger law],
equation (2.2) below, these findings are generic. Hence the
possible foundational character of the Local Time Scheme not only
in regard of open quantum systems theory but also regarding
certain cosmological issues.

In Section 2 and with the aid of Appendix A, we briefly overview
and discuss certain subtle points in Local Time Scheme. On this
basis we investigate mathematical characteristics of the LTS
dynamical map, in Section 3 for a closed many-particle system and
in Section 4 for an open system in contact with a many-particle
environment. In Section 5 we provide quantitative criteria for
Markovian dynamics for a closed many-particle system. In Section 6
we compare the obtained results with the related counterparts from
the standard open systems theory. Section 7 is discussion, where
we place an emphasis on certain cosmological issues. Section 8 is
conclusion.

\bigskip

{\bf 2. Outlines of Local Time Scheme}

\bigskip

Formally, Local Time Scheme introduces a variation of the standard
unitary dynamics of a closed quantum system.

The standard unitary dynamics

\begin{equation}
\vert \Psi(t_{\circ})\rangle =  U(t_{\circ}) \vert \Psi(0)\rangle
\end{equation}

\noindent is exchanged by a statistically weighted the 'final'
instant of time $t_{\circ}$:

\begin{equation}
\sigma(t_{\circ}) = \int_{t_{\circ}-\Delta t}^{t_{\circ} + \Delta
t} dt \rho(t) \vert \Psi(t)\rangle\langle\Psi(t)\vert =
\int_{t_{\circ}-\Delta t}^{t_{\circ} + \Delta t} dt \rho(t) U(t)
\vert \Psi(0)\rangle\langle\Psi(0)\vert U^{\dag}(t).
\end{equation}

\noindent We do not see any alternative to equation (2.2) within
the {\it minimal} extension [1] of the standard theory, which
adopts  equation (2.1).

The origin of equation (2.2), which is the fundamental dynamical
law in LTS, is an idea proposed within the quantum many-body
scattering theory. While investigating the problem of asymptotic
completeness in the many-body scattering theory, Hitoshi Kitada
[2] noticed that unitary dynamics allows for an operational {\it
definition} of time. The standard 'time instant' becomes a
(one-dimensional) classical parameter that is determined by the
unitary (Schr\" odinger) dynamics of a closed (or approximately
closed) quantum system in the asymptotic limit; such systems are
routinely assumed in  quantum decoherence and measurement as well
as in the open systems theory [15-18]. Hence the possibility that
different Hamiltonians determine different times, each local time
being a characteristic of a 'local' (approximately closed) system.
LTS is a kind of 'multi-time' theory that does not suffer from
some known drawbacks of fixed local times [1,19].

The fact that local time is defined in the {\it asymptotic} limit
[2] leads to equation (2.2), since the {\it finite} value
$t_{\circ}$ for a single local system is not uniquely defined but
only within some interval $\Delta t$. In order not to approximate
equation (2.2) by equation (2.1), the interval $\Delta t$ for the
continuous parameter $t_{\circ}$ should not be arbitrarily small.
Hence the requirement $t_{\circ} \gg \Delta t$ implies that
$t_{\circ}$ cannot be arbitrarily small either, while the formal
limit $t_{\circ} \to \infty$ should be allowed.

On the other hand, for some values of $\Delta t$, certain states
$\vert \Psi(t)\rangle$ in equation (2.2) can be mutually
orthogonal [20, 21]. As this seems to contradict equation (2.1), a
proper upper bound, $\tau_{min}$, for $\Delta t$ has been
recognized that yields the constraint $\tau_{min}
> \Delta t$. The choice of a Gaussian probability density $\rho(t)
= \sqrt{\lambda/\pi} \exp(-\lambda(t-t_{\circ})^2)$ gives rise to
the constraint $\tau_{min} > \Delta t > \lambda^{-1/2}$. The
condition $\tau_{min} = max\{\pi\hbar/2\Delta H,
\pi\hbar/2(\langle H\rangle - E_g)\}$, where $\Delta H$ is the
standard deviation and $E_g$ stands for the Hamiltonian ground
energy, gives rise to the estimation $\lambda > C^2$, where $C$
represents the energy scale of the system's Hamiltonian. It is the
basic characteristic of LTS: there is energy conservation
 for the state equation (2.2) and hence time independence of both
$\Delta H$ and $\langle H\rangle$. Nonorthogonality of states
$\vert \psi(t)\rangle$ in equation (2.2) makes the local time
instants from the interval $[t_{\circ}-\Delta t, t_{\circ}+\Delta
t]$ mutually indistinguishable. Hence  Local Time as a hidden
classical parameter of the  system's dynamics.

Within Local Time Scheme, a single system that is subjected to a
local time $t_1$ may interact with another single system that is
subjected to a local time $t_2$. Assume that their {\it
sufficiently strong} interaction starts in an instant of time
$t_{1\circ}$ for the first, and in an instant $t_{2\circ}$ of
local time for the second system.  The start of the interaction
defines a new closed system, $1+2$, whose Hamiltonian determines
the composite system's local time, with the new initial 'instant
of time' $t=0$ for the combined dynamics. For the previously
non-interacting systems undergoing independent unitary dynamics
subjected to their independent local times, the tensor-product
initial state is $\vert\psi(t_{1\circ})\rangle_1 \vert
\phi(t_{2\circ})\rangle_2 \equiv \vert \Psi(t=0)\rangle$, which
appears in equation (2.2). This description may raise the
following three concerns regarding consistency and coherence of
LTS. First, the meaning of 'sufficiently strong' interaction (that
 underlies the meaning of 'approximately closed' ('local') system)
is not obvious. Second, in realistic situations, a smooth
dynamical transition from 'weak' to 'sufficiently strong'
interaction is expected and should be properly described. Finally,
provided the answers to these questions, it is  natural to ask
about uniqueness of the 'initial instant' $t=0$ for the combined
system. Putting $t_{\circ}=0$ in equation (2.2) reveals
non-uniquely defined 'initial time instant' for an ensemble of
$1+2$ systems. On the other hand, the local time instants
$t_{i\circ }, i=1,2$, can locally satisfy $t_{i\circ} \to \infty$
as per equation (2.2)--i.e. from the point of view of the local
$1$ and $2$ systems, these instants are not special and are
subject to the time uncertainty equation (2.2). In Ref. [1] we
analyzed the standard scenario of unique instant of time in an
ensemble presented by equation (2.1) and hence adopted equation
(2.2) without variations. The first two questions are addressed in
Appendix A. Non-unique initial instant of time is the subject of
lemma 3.3 below.

\bigskip

{\bf 3. Closed system dynamics}

\bigskip

Dynamics introduced by equation (2.2) is linear. Therefore its
generalization for  mixed states

\begin{equation}
\sigma(t_{\circ}) = \int_{t_{\circ}-\Delta t}^{t_{\circ} + \Delta
t} dt \rho(t)  U(t) \sigma(0)  U^{\dag}(t),
\end{equation}

\noindent with a dynamical map $\mathcal{E}$

\begin{equation}
\sigma(t_{\circ}) = \mathcal{E}_{(t_{\circ}, 0)} [\sigma(0)]
\end{equation}

\noindent that we are interested in.

Equation (3.1) is of the form of the so-called 'random-unitary'
maps (RUM) [22]. However, as distinct from the standard RUMs, of
the general form $\sum_i p_i U_i(t) \sigma(0) U^{\dag}_i(t),
\sum_i p_i = 1, \forall{t}$, and unitary $U_i, \forall{i}$, the
map equation (3.1) varies the time instant(s).

Inclusion of the unitary operator spectral form, $U(t) = \sum_n
\exp(-\imath tE_n/\hbar) P_n$, where $E_n$ and $P_n$ stand for the
 Hamiltonian eigenvalues and orthogonal eigenprojectors,
gives for equation (3.1):

\begin{equation}
\sigma(t_{\circ}) = \sum_{m,n} \exp\left({-\imath t_{\circ}(E_m -
E_n)\over \hbar}\right) \exp\left({-(E_m - E_n)^2\over
4\lambda\hbar^2}\right) P_m \sigma(0) P_n.
\end{equation}

\noindent Integration in equation (3.1) gives equation (3.3) via
the use of the following Gaussian integral:
$\int_{-\infty}^{\infty} \exp(-ax^2/2 + \imath Jx) dx =
(2\pi/a)^{1/2}\times$ $\exp(-J^2/2a)$, where $a
> 0$ and $J$ are real numbers with $J$ being conjugate variable of
$x$.

From the mutually equivalent equations (3.1)-(3.3) directly
follow: (i) the map is linear, positive and trace preserving; (ii)
the map is unital, i.e. $\mathcal{E}[I] = I$; (iii) the r.h.s. of
equation (3.1) is of the so-called Kraus form  that guarantees
that the map is completely positive (CP). Due to the non-standard
character of the map to be emphasized below, we give an
alternative proof of complete positivity that employs the
Jamiolkowski criterion [23]. According to this criterion, a map
$\mathcal{E}$ is CP if and only if the extended map
$\mathcal{I}\otimes \mathcal{E}$ is positive while acting on a
maximally entangled state $\vert \psi\rangle$ for an extended
 system:

\begin{equation}
\langle \phi \vert \left(\mathcal{I}\otimes\mathcal{E}[\vert
\psi\rangle\langle\psi\vert]\right)\vert \phi\rangle \ge 0,
\forall{\vert \phi\rangle}.
\end{equation}

By putting $\vert \psi\rangle = \sum_i \vert i\rangle\vert
i\rangle/d$ and arbitrary state $\vert \phi \rangle = \sum_{j,k}
c_{jk}\vert j\rangle\vert k\rangle$ into equation (3.4), the
criterion reads

\begin{equation}
{1\over d} \sum_{i,j,l,l'} c^{\ast}_{il} c_{jl'} \langle l\vert
\left(\mathcal{E}[\vert i\rangle\langle j\vert]\right)\vert
l'\rangle \ge 0.
\end{equation}

For the map equation (3.1), the criterion equation (3.5) easily
gives:

\begin{equation}
\int_{t_{\circ} -\Delta t}^{t_{\circ} + \Delta t} dt \rho(t)
\left\vert \sum_{i,l} c^{\ast}_{il} \langle l\vert U(t)\vert
i\rangle\right\vert^2
> 0.
\end{equation}

However,  the map  is rather non-standard as we are going to
demonstrate. For the initial $t=0$ as well as for arbitrary value
of the time parameter $t$, the map satisfies:

\begin{equation}
\mathcal{E}_{(t,t)} \neq \mathcal{I}, \forall{t}
\end{equation}

\noindent where the $\mathcal{I}$ stands for the identity (unity)
map. Putting $t_{\circ} = 0$ in equation (3.1) does not return, as
apparently it should, the initial state $\sigma (0)$.
 Hence, as
 distinct from the standard theory, equation (2.1), in which
$t_{\circ} \in (-\infty, \infty)$, in Local Time Scheme, equation
(2.2), $t_{\circ} \in (0, \infty)$.

The in-equality (3.7) is striking: the map describes a {\it
non-differentiable process}, that is  incapable of deriving a
differential form for the  equation  (3.3). Actually, derivation
of a differential form of the open system dynamical law {\it
assumes} [15-17] non-validity of equation (3.7), i.e. validity of
$\mathcal{E}_{(t,t)} = \mathcal{I}, \forall{t}$, in order to have
a mathematically properly defined 'generator' of evolution [often
denoted $\mathcal{L}_t$]. Therefore we proceed with the analysis
of the 'integral' form of the dynamical law equation (3.3).

\bigskip

{\bf 3.1 Non-divisibility of the map}

\bigskip

\noindent {\bf Lemma 3.1 } {\it The map}  (3.1)  {\it can be
combined with the unitary dynamics}:

\begin{equation}
\mathcal{E}_{(t_{\circ}, 0)} [\sigma(0)] =
\mathcal{U}_{(t_{\circ}, t'')}\left[\mathcal{E}_{(t'', t')}
\left[\mathcal{U}_{(t',0)}[\sigma(0)]\right]\right], \quad
t_{\circ} \ge t'' \ge t' > 0,
\end{equation}

\noindent {\it where} $\mathcal{U}$ {\it denotes the unitary
dynamics equation} (2.1).

\noindent{\it Proof}. From equation (3.3) we can directly write:

\begin{eqnarray}
&\nonumber& \sigma(t_{\circ}) =
\mathcal{E}_{(t_{\circ},t')}[\mathcal{U}_{(t',0)}[\sigma(0)]] =
\\&& \sum_{m,n} \exp\left({-\imath (t_{\circ}-t')(E_m - E_n)\over
\hbar}\right) \exp\left({-(E_m - E_n)^2\over
4\lambda\hbar^2}\right) P_m \sigma(t') P_n.
\end{eqnarray}

\noindent  where $\sigma(t')$ refers  to the unitary evolution as
well as

\begin{equation}
\sigma(t_{\circ}) =
\mathcal{U}_{(t_{\circ},t')}[\mathcal{E}_{(t',0)}[\sigma(0)]] =
\sum_{m,n} \exp\left({-\imath (t_{\circ}-t')(E_m - E_n)\over
\hbar}\right) P_m \sigma(t') P_n
\end{equation}

\noindent with $\sigma(t')$  given by equation (3.3).

Now equation (3.9) gives:

\begin{eqnarray}
&\nonumber& \sigma(t_{\circ}) =  \sum_{m,n} \exp\left({-\imath
(t_{\circ}-t')(E_m - E_n)\over \hbar}\right) \exp\left({-(E_m -
E_n)^2\over 4\lambda\hbar^2}\right)
\\&& P_m \left[\sum_{p,q}
\exp(-\imath t' (E_p - E_q)/\hbar) P_p \sigma(0) P_q \right] P_n.
\end{eqnarray}

From equation (3.10):

\begin{eqnarray}
&\nonumber& \sigma(t_{\circ}) =
\mathcal{U}_{(t_{\circ},t')}[\mathcal{E}_{(t',0)}[\sigma(0)]] =
\sum_{m,n} \exp\left({-\imath (t_{\circ}-t')(E_m - E_n)\over
\hbar}\right)
\\&& P_m \left[ \sum_{p,q} \exp\left({-\imath t'(E_p - E_q)\over
\hbar}\right) \exp\left({-(E_p - E_q)^2\over
4\lambda\hbar^2}\right) P_p \sigma(0) P_q \right]
 P_n.
\end{eqnarray}

Orthogonality of the projectors, $P_m P_n = \delta_{mn} P_m$, with
the 'Kronecker delta', $\delta_{mn}$, gives equation (3.3) for
both equations  (3.11) and (3.12). Combination of equations (3.11)
and (3.12) completes the proof. \hfill Q.E.D.

Lemma 3.1 reveals that the time uncertainty, originally linked
with $t_{\circ}$ in equation (3.1), formally applies to every {\it
single} value $t \in (0, t_{\circ}]$ but not to any time interval.
Thereby we learn that the LTS time-uncertainty equation (2.2)
cannot be interpreted as or be reducible to action of a 'quantum
channel' [3,22].

\noindent {\bf Lemma 3.2} {\it The map cannot be divided according
to the law}  (3.2):

\begin{equation}
\mathcal{E}_{(t_{\circ}, 0)} [\sigma(0)] \neq
\mathcal{E}_{(t_{\circ}, t')} \left[ \mathcal{E}_{(t', 0)}
[\sigma(0)]\right], \quad t_{\circ} \ge t' > 0.
\end{equation}

\noindent{\it Proof}. Due to equation (3.3), the r.h.s. of
equation (3.13) reads:

\begin{eqnarray}
&\nonumber&  \sigma(t_{\circ}) = \sum_{m,n} \exp\left({-\imath
(t_{\circ}-t')(E_m - E_n)\over \hbar}\right) \exp\left({-(E_m -
E_n)^2\over 4\lambda\hbar^2}\right)
\\&& \nonumber P_m\left[
\sum_{p,q} \exp\left({-\imath t'(E_p - E_q)\over \hbar}\right)
\exp\left({-(E_p - E_q)^2\over 4\lambda\hbar^2}\right) P_p
\sigma(0) P_q. \right] P_n\\&&  = \sum_{m,n} \exp\left({-\imath
t_{\circ}(E_m - E_n)\over \hbar}\right) \exp\left({-2(E_m -
E_n)^2\over 4\lambda\hbar^2}\right) P_m \sigma(0) P_n.
\end{eqnarray}

Now decomposition of the interval $(0, t_{\circ}]$ into $k$
subintervals, with the aid of equation (3.14), gives instead of
equation (3.3):

\begin{equation}
\sum_{m,n} \exp\left({-\imath t_{\circ}(E_m - E_n)\over
\hbar}\right) \exp\left({-k(E_m - E_n)^2\over
4\lambda\hbar^2}\right) P_m \sigma(0) P_n \to  \sum_{m} P_m
\sigma(0) P_m
\end{equation}

\noindent as $k \to \infty$ that completes the proof. \hfill
Q.E.D.

Lemma 3.2 exhibits non-validity of the assumptions of Def.1.1 and
hence reveals non-Markovian character of the map. In the spirit of
the standard wisdom [16,17], the total map,
$\mathcal{E}_{(t_{\circ}, 0)}$, and the initial map,
$\mathcal{E}_{(t', 0)}$, [that are given by equation (3.1)] are
completely positive. However, according to Lemma 3.2, the
'intermediate' map, $\mathcal{E}_{(t_{\circ}, t')}$, is not and
therefore equation (3.1), i.e. equation (3.3), does not apply to
$\mathcal{E}_{(t_{\circ}, t')}$.

In addition, we note that equation (3.3) can be transformed to a
Kraus form that is alternative to the integral form of equation
(3.1). Complete positivity $\langle \chi\vert
\sigma(t_{\circ})\vert \chi\rangle\ge 0$, $\forall{\vert
\chi\rangle}, t_{\circ} $, cf. equation (3.6), implies  the matrix
$\mathcal{A}=(\exp({-(E_m - E_n)^2/ 4\lambda\hbar^2}))$ is
positive semi-definite. Then diagonalization of the $\mathcal{A}$
matrix by a unitary $\mathcal{U} = (u_{km})$ matrix gives $A_{mn}
= \sum_k \gamma_k u^{\ast}_{km} u_{kn}$ with the eigenvalues
$\gamma_k \ge 0$, $\forall{k}$, while $A_{mm}=1, \forall{m}$.
Therefore

\begin{equation}
\sigma(t_{\circ}) = \sum_k K_k(t_{\circ}) \sigma(0)
K_k^{\dag}(t_{\circ}),
\end{equation}

\noindent where the Kraus operators $K_k(t_{\circ}) =
\sqrt{\gamma_k} \sum_m u^{\ast}_{km}\exp(-\imath
t_{\circ}E_m/\hbar) P_m$ and the equality $\sum_k
K^{\dag}_k(t_{\circ})$ $K_k(t_{\circ}) = I$ is satisfied. However,
due to equation (3.7), existence of a Lindblad differential form
of equation (3.16) is excluded.

\noindent {\bf Lemma 3.3} {\it Nonuniqueness of the initial} $t=0$
{\it in equation} (3.1) {\it does not change the character or the
form of the map}.

\noindent {\it Proof}. A change of local time occurs  due to
sufficiently strong  interaction of two systems not being
subjected to the same local time. Then uncertainty of the initial
'instant' $t=0$ regards uncertainty of duration of the
approximately independent dynamics of the two systems. Then
equation (3.3) gives

\begin{equation}
\sigma(t_{\circ}) = \sum_{m,n} \exp\left({-\imath t_{\circ}(E_m -
E_n)\over \hbar}\right) \exp\left({-(E_m - E_n)^2\over
4\lambda\hbar^2}\right) P_m \sigma'(0) P_n.
\end{equation}

\noindent where

\begin{equation}
\sigma'(0) = \int_{-\delta t}^{\delta t} \rho'(t) U'(t) \sigma(0)
U'^{\dag}(t) dt.
\end{equation}

For  non-interacting, or weakly interacting systems the unitary
operator $U'(t) = \sum_{p}$ $\exp(-\imath tE'_p/\hbar) p_p $ and
$\sigma(0) = \vert \Psi(0)\rangle\langle\Psi(0)\vert$, while
$\vert \Psi(0)\rangle = \vert\psi(t_{1\circ})\rangle_1 \vert
\phi(t_{2\circ})\rangle_2$ and $\rho'(t) = \sqrt{\lambda'/\pi}$
$\exp(-\lambda' t^2)$, equation (3.18) reads [e.g. by putting
$t_{\circ}=0$ in equation (3.3)]:

\begin{equation}
\sigma'(0) = \sum_{p,q} \exp\left({-(E'_p - E'_q)^2 \over
4\lambda'\hbar^2}\right) p_p \sigma(0)p_q.
\end{equation}

The two uncertainties, $\Delta t$ and $\delta t$, determine the
two Gaussian factors, $\lambda$ and $\lambda'$, respectively. In
order to comply with equation (2.1), $\delta t$ should not exceed
$\Delta t$. Hence, cf. Section 2, $\lambda'$ cannot be smaller
than $\lambda$; we could have introduced two $\lambda'$s for every
subsystem separately but this would change nothing. If we
introduce the fixed (cf. Appendix A) energy scales $c$ and $C$ for
the two Hamiltonians, the condition that interaction dominates the
total system's dynamics  yields $c \ll C$ (in the units $\hbar =
1$). Bearing in mind (Section 2) that $\lambda' >\lambda
> C^2$, the exponential factors in equation (3.19) can all be estimated
$\exp(- c^2/4C^2) \approx 1$, and hence $\sigma'(0) \approx
\sigma(0)$. \hfill   Q.E.D.

Lemma 3.3 quantitatively, i.e. via the condition $c/C \ll 1$,
removes the ambiguity regarding $t=0$ in Local Time Scheme: for
given $c$, strong interaction ($C\gg c$) gives rise to practically
indistinguishable dynamics equation (3.3) and equation (3.17).
Interestingly, there is a question of handling the same for the
standard theory, equation (2.1). Indeed, initial preparation of a
quantum ensemble may be not  pure as generally assumed and
described by equation (2.1). Then equation (2.2) with
$t_{\circ}=0$ applies to the standard theory of unique time and
hence produces an observable, although probably weak, dissent with
equation (2.1).  A plausible answer might be that $\delta t$
should be so small as to provide $U(\delta t) = I + O(\delta t)$,
which gives equation (2.1) approximately correct. Needless to say,
this option also applies to our considerations again leading to
$\sigma'(0) \approx \sigma(0)$. Hence LTS appears to be more
robust to variations of the 'initial instant' than the standard
theory of unique time.

\bigskip

{\bf 3.2 An approximation of the map}

\bigskip

Numerical values of the real Gaussian factors in equation (3.3)
depend on the energy-spectrum $\{E_m\}$ and also on the initial
state $\vert \Psi(0)\rangle$. Consider a coarse graining of the
set of energies: every value $E_m$ is assigned a set of
numerically (or operationally) close values $E_{\nu_m}$. The
standard procedure would be to introduce a new set of energy
eigenvalues by setting $E_m = E_{\nu_m}$ and the related
eigenprojectors that redefine the system's Hamiltonian, and then
to start over from equation (2.1), i.e. from equation (3.1).
However, we will proceed in the following, more flexible
operational fashion that is closer [but not identical] in spirit
to [24].

We directly and independently adapt the Gaussian terms appearing
in equation (3.3) via the numerical estimates:

\begin{equation}
 \exp\left( -
{(E_{k} - E_{k'})^2 \over 4\lambda\hbar^2}\right) \ll 1, \quad
 k\in\{m, \nu_m\}, k'\in\{m',\nu'_{m'}\}, \forall{m\neq m'}.
\end{equation}

Being interested in large values of $t_{\circ}$, we  set
$E_{\nu_m} - E_{\nu'_m} \approx \delta_m > 0, \forall {m,\nu,
\nu'}$ and assume $\exp\left( - {\delta_m^2 /
4\lambda\hbar^2}\right) \approx 1, \forall{m}$, that can follow
from the numerical values or determine the measurement errors, and
obtain:

\begin{equation}
\sigma(t_{\circ}) \approx \sum_m P_m \sigma(0) P_m +
 \sum_m
\exp(-\imath t_{\circ}\delta_m/\hbar) P_m \sigma(0) \Pi^{(m)} +
h.c.
\end{equation}

Numerical details behind equation (3.21) can be found in Section
5. Equation  (3.20) is typically not applicable to the
few-particle systems, which are already known [1] to bear high
quantum coherence, i.e. approximately to be dynamically described
by equation (2.1).

In equation (3.21), for typically non-orthogonal but commuting
projectors $\Pi^{(m)} = \sum_{\nu_m}P_{\nu_m}$ uniquely defined by
the chosen coarse graining: (i) $P_m \Pi^{(m)} = 0, \forall{m}$,
while the commutator $[P_m, \Pi^{(m')}]=0, \forall{m,m'}$; (ii) if
$P_m \Pi^{(m')} \neq 0$ for some $m$ and $m'$, then $P_{m'}
\Pi^{(m)} = 0$ for the same pair of indices $m$ and $m'$; (iii) it
is allowed that $\delta_m = \delta_{m'}$ for some $m\neq m'$.

The map equation (3.21) is linear, unital, trace preserving and
applies equation (3.7) as well as  lemma 3.1.

For the $d$-independent coarse graining, i.e. for the
coarse-graining parameter $g=max\{tr \Pi^{(m)}\}$ such that
$\lim_{d\to\infty} g = g$:

\noindent {\bf Lemma 3.4} {\it The approximate map } (3.21) {\it
is completely positive for almost all large values of}
$t_{\circ}$, i.e.  {\it formally, the map is completely positive
in the limit} $t_{\circ} \to \infty$.

\noindent {\it Proof}. We start from the lhs of  equation (3.5),
which for  equation (3.21) gives:

\begin{eqnarray}
&\nonumber&  {1\over d} \sum_m  \left\vert \sum_{i,l}
c^{\ast}_{il} \langle l\vert P_m\vert i\rangle \right\vert^2 +
\\&& {1\over d} \sum_m
\exp(-\imath t_{\circ}\delta_m/\hbar) \left( \sum_{i,l}
c^{\ast}_{il} \langle l\vert P_m\vert i\rangle \right) \left(
\sum_{j,l'} c_{jl'} \langle j\vert \Pi^{(m)}\vert l'\rangle
\right) + c.c.
\end{eqnarray}

Since the basis $\vert i\rangle$ appearing in equation (3.5) can
be chosen so as $\langle l\vert P_m\vert i\rangle = 0,
\forall{l\neq i}$ and $\langle j\vert \Pi^{(m)}\vert l'\rangle =
0, \forall{j\neq l'}$,
 the second term in equation (3.22):

\begin{equation}
{1\over d} \sum_m \left(\sum_{i=1}^{g_m} c^{\ast}_{ii}\right)
\left( \sum_{\nu_i=1}^{g^{(m)}} c_{\nu_i \nu_i}\right),
\end{equation}

\noindent where $g_m = tr P_m$ and $g^{(m)} = tr \Pi^{(m)}$. As we
are interested in the largest possible value for the sum equation
(3.23), we introduce the real $c_{ii} = p_i \ge 0$ and
$c_{\nu_i\nu_i} = p_{\nu_i} \ge 0$; the largest value follows for
$\sum_{i=1}^d p_i^2 =1$--that pertains to the very special
[normalized] states $\vert \phi\rangle = \sum_i p_i \vert
i\rangle\vert i\rangle$ in equation (3.4). For all other choices
of the complex $c_{ii}$ ($c_{\nu_i\nu_i}$) the second (third) term
in equation (3.22) is modulo smaller.

Regarding equation (3.23):

\begin{equation}
  \sum_{m=1}^N \chi_m \equiv  \sum_{m=1}^N
\left(\sum_{i=1}^{g_m} p_i\right) \left( \sum_{\nu_i=1}^{g^{(m)}}
p_{\nu_i}\right) \le g_{\max} g  N, \quad g_{\max} = \max\{ g_m,
m=1,2,...,N\}.
\end{equation}

With the notation $C \equiv  \sum_{m=1}^N  \left( \sum_{i=1}^{g_m}
p_i \right)^2 > 0$ we are interested to prove

\begin{equation}
1 + {2g g_{max} N\over C}\sum_{m=1}^N \cos(\delta_m t_{\circ})
{\chi_m\over g g_{max} N} \ge 0
\end{equation}

\noindent for almost all large values of $t_{\circ}$ and for
sufficiently large $N$.

From equation (3.24) follows $\sum_{m=1}^N \chi_m/g g_{max} N \le
1$ and hence the sum in equation (3.25) is an almost periodic
function [25-27]. For sufficiently long time interval $[t, t+T]$,
and a large number $N$ (that increases with the increase of the
number of particles in the system [27]) such that $t_{\circ} \in
[t, t+T]$ typically [18,27]: (a) the time average
$\lim_{T\to\infty} \langle \sum_m \cos(\delta_m t_{\circ})
{\chi_m\over g g_{max} N}\rangle_T$ $= 0$, and (b) the standard
deviation $\lim_{T\to\infty} \langle \vert \sum_m$  $
\cos(\delta_m t_{\circ}) {\chi_m\over g g_{max} N}
\vert^2\rangle_T = 0$. On the other hand $C = \sum_{m=1}^N (g_m
\bar p_m)^2$, where $\bar p_m$ is the average value for the set
$\{p_{i}, i = 1,2,...,g_m\}$ and so $C \ge g_{\min}^2 \sum_{m=1}^N
{\bar p_m}^2$. So the term $2g g_{\max} N/ C \le 2g_{\max} g
/\sum_{m=1}^N {\bar p_m}^2$. Noting that $g_{max} < g$ and the
fact that $\sum_{m=1}^N {\bar p_m}^2/N = (\Delta \bar p)^2 +
\langle \bar p\rangle^2$ never decreases with the increase of $N$,
the term $2g g_{max} N/ C$ does not increase with the increase of
$N$; $\Delta \bar p$ and $\langle \bar p\rangle$ are the standard
deviation and the average value for the set $\{\bar p_m\}$.
Bearing in mind that the probability distribution for the sum in
equation (3.25) goes to the Dirac delta-function as $N\to\infty$
[27], the probability that the sum over $m$ in equation (3.25)
takes a value less than $-C/2g g_{max} N$ approaches zero in the
limit $N\to\infty$ [27]. Therefore we proved  equation (3.25).
Multiplying equation (3.25) by $C/d$, for a not-very-large coarse
graining constant $g$, from equation (3.22):

\begin{equation}
{1\over d} \sum_{m=1}^N  \left( \sum_{i=1}^{g_m} p_i \right)^2 +
2g\sum_{m=1}^N \cos(\delta_m t_{\circ}) {\chi_m\over gd} \ge 0,
\end{equation}

\noindent that is, for most of  large values of $t_{\circ}$ and
for sufficiently large $N$ [large number of particles in the
system] the approximate map equation (3.21) is completely
positive. \hfill Q.E.D.

 Lemma 3.4
exhibits {\it dynamical} emergence of complete positivity of the
approximate map equation (3.21).  The larger value of $g$ the
larger the 'minimum' value of $t_{\circ}$ for which equation
(3.26) is fulfilled. For exceedingly large $g$, the minimum value
of $t_{\circ}$ may be very large and also equation (3.26) less
likely to be satisfied for some instants $t > t_{\circ}$.

Regarding divisibility of the map equation (3.21):

\noindent {\bf Lemma 3.5} {\it The approximate map }
$\mathcal{E}^{(app)}$ {\it equation } (3.21) {\it is divisible}:

\begin{equation}
\mathcal{E}^{(app)}_{(t_{\circ},0)}[\sigma(0)] =
\mathcal{E}^{(app)}_{(t_{\circ},t')}\left[\mathcal{E}^{(app)}_{(t',0)}[\sigma(0)]\right],
\quad t_{\circ} \ge t' > 0.
\end{equation}

\noindent {\it Proof}. Equation  (3.21) gives:

\begin{equation}
\sigma(t_{\circ}) \approx \sum_m P_m \sigma(t') P_m + \sum_m
\exp(-\imath (t_{\circ}-t')\delta_m/\hbar) P_m \sigma(t')
\Pi^{(m)} + h.c.
\end{equation}

Substituting equation (3.21) with $t'$ instead of $t_{\circ}$ into
equation (3.28), due to the above point (i) equation (3.28) reads:

\begin{eqnarray}
&\nonumber&  \sigma(t_{\circ}) \approx \sum_m P_m \sigma(0) P_m +
\sum_m \exp(-\imath t_{\circ}\delta_m/\hbar) P_m \sigma(0)
\Pi^{(m)} +\\&& \sum_{m,n} \exp(-\imath
(t_{\circ}-t')\delta_m/\hbar) \exp(\imath t'\delta_n/\hbar) P_m
\Pi^{(n)} \sigma(0) P_n \Pi^{(m)} + h.c.
\end{eqnarray}

Due to the above point (ii), the last  term in equation (3.29)
equals zero, thus equation (3.29) taking the form of equation
(3.21). \hfill Q.E.D.

Now it is obvious that lemma 3.4 and lemma 3.5 give rise to the
conclusion: as distinct from the exact map equation (3.3), the
approximate map equation (3.21) {\it dynamically acquires
Markovianity}, Def.1.1. Due to  Lemma 3.4, Markovian character of
the map equation (3.21) requires the lower time bound, i.e. coarse
graining of the time interval $(0,t_{\circ}]$--thus resembling the
so-called Born approximation in the standard theory [15-17]. In
accordance with Section 2, the initial state for the $S+E$ system
is assumed to be tensor-product.

\bigskip

{\bf 4. Open system dynamics}

\bigskip

We are interested in a bipartite decomposition of the total system
of Section 3 that consists of two subsystems, $S$ and $E$, in
which  interaction between $S$ and $E$ dominates the total
system's dynamics, which is the situation described at the end of
Section 2.

Then the unitary operator $U(t) \approx \exp(-\imath t
H_{int}/\hbar)$ and we consider the 'pure decoherence' interaction
given by the 'separable' spectral form [27,28]:

\begin{equation}
 H_{int} = \sum_{\alpha, \beta} E_{\alpha\beta} P_{\alpha}
\otimes \Pi_{\beta},
\end{equation}

\noindent where the orthogonal projectors $P_{\alpha}$ refer to
the $S$ system,  the projectors $\Pi_{\beta}$ refer to the $E$
system, while the interaction eigenvalues $E_{\alpha\beta}$ are
all real such that $E_{\alpha\beta} = E_{\gamma\delta}$ if and
only if $\alpha =\beta$ and $\gamma = \delta$. In accordance with
Section 2, the initial state for the $S+E$ system is
tensor-product that gives rise to the composite system's state of
the form of equation (3.3).

\bigskip

 {\bf 4.1 The exact map}

 \bigskip

With the use of equation (4.1), after some algebra from equation
(3.3) follows:

\begin{eqnarray}
&\nonumber&  \rho_S(t_{\circ}) = tr_E \sigma(t_{\circ}) =
\sum_{\alpha,\gamma}P_{\alpha} \rho_S(0) P_{\gamma} \times
\\&& \nonumber
\sum_{\beta}\exp \left(- \imath t_{\circ} {E_{\alpha\beta} -
E_{\gamma\beta}\over \hbar}\right) \exp \left(-{(E_{\alpha\beta} -
E_{\gamma\beta})^2\over 4\lambda\hbar^2}\right) (tr_E
\Pi_{\beta}\rho_E(0))\\&& \equiv \sum_{\alpha,\gamma}
B_{\alpha\gamma}(t_{\circ}) P_{\alpha} \rho_S(0) P_{\gamma}.
\end{eqnarray}

Similarly, for the successive time intervals
$(0,t'],[t',t_{\circ}]$ and for $\sigma(t')$ given by equation
(3.3):

\begin{eqnarray}
&\nonumber&  \rho_S(t_{\circ}) = tr_E \sigma(t_{\circ}) =
\sum_{\alpha,\beta,\gamma,\delta}
   P_{\alpha} \left[
tr_E \Pi_{\beta} \sigma(t')      \Pi_{\delta}\right] P_{\gamma}
\times
\\&&\nonumber \exp \left(- \imath
(t_{\circ}-t') {E_{\alpha\beta} - E_{\gamma\delta}\over
\hbar}\right) \exp \left(-{(E_{\alpha\beta} -
E_{\gamma\delta})^2\over 4\lambda\hbar^2}\right)\\&& \nonumber =
\sum_{\alpha,\gamma} P_{\alpha} \rho_S(0) P_{\gamma} \times
\\&& \nonumber\sum_{\beta}\exp \left(- \imath
t_{\circ} {E_{\alpha\beta} - E_{\gamma\beta}\over \hbar}\right)
\exp \left(-2{(E_{\alpha\beta} - E_{\gamma\beta})^2\over
4\lambda\hbar^2}\right) (tr_E \Pi_{\beta}\rho_E(0))
\\&& \equiv \sum_{\alpha,\gamma}
B'_{\alpha\gamma}(t_{\circ}) P_{\alpha} \rho_S(0) P_{\gamma}.
\end{eqnarray}

From equation (4.2)  it easily follows that the map
 is linear, positive, trace preserving and unital. Like
the exact map, Section 3, the map fulfills equation (3.7). As the
matrix $(B_{\alpha\gamma}(t_{\circ}))$ is positive semi-definite
(cf. equation (3.16)), the map is completely positive for every
$t_{\circ}
> 0$.

On the other hand, {\it prima facie}  equation (4.3) exhibits
non-divisibility of the
 map: decomposition of the interval
$(0,t_{\circ}]$ into $k$ subintervals leads, in analogy with
equation (3.15), to

\begin{eqnarray}
&\nonumber&  \sum_{\alpha,\gamma} P_{\alpha} \rho_S(0) P_{\gamma}
\times
\\&&\nonumber\sum_{\beta}\exp \left(- \imath
t_{\circ} {E_{\alpha\beta} - E_{\gamma\beta}\over \hbar}\right)
\exp \left(-k{(E_{\alpha\beta} - E_{\gamma\beta})^2\over
4\lambda\hbar^2}\right) (tr_E \Pi_{\beta}\rho_E(0))\\&& \to
\sum_{\alpha} P_{\alpha} \rho_S(0) P_{\alpha}
\end{eqnarray}

\noindent as $k \to \infty$. However, the map dynamically acquires
divisibility as stated by the following

\noindent {\bf Lemma 4.1}   {\it The map}  (4.2) {\it has unique
steady state and  acquires divisibility for almost all large
values of} $t_{\circ}$, i.e. {\it formally, the map is divisible
in the limit} $t_{\circ} \to \infty$:

\begin{equation}
\lim_{t_{\circ} \to \infty} \rho_S(t_{\circ}) = \sum_{\alpha}
P_{\alpha} \rho_S(0) P_{\alpha}.
\end{equation}

\noindent {\it Proof}. For equation (4.2)

\begin{equation}
\rho_{S}(t_{\circ}) = \sum_{\alpha,\gamma}
B_{\alpha\gamma}(t_{\circ})  P_{\alpha} \rho_S(0) P_{\gamma}
\end{equation}

\noindent where

\begin{equation}
B_{\alpha\gamma}(t_{\circ}) = \zeta_{\alpha\gamma} \sum_{\beta}
p_{\beta}^{(\alpha\gamma)} \exp \left(- \imath t_{\circ}
{E_{\alpha\beta} - E_{\gamma\beta}\over \hbar}\right)
\end{equation}

\noindent while the real $p_{\beta}^{(\alpha\gamma)}= (tr_E
\Pi_{\beta}\rho_E(0))\exp \left(-{(E_{\alpha\beta} -
E_{\gamma\beta})^2\over
4\lambda\hbar^2}\right)/\zeta_{\alpha\gamma} > 0$,
$\sum_{\beta}p_{\beta}^{(\alpha\gamma)}=1$ and
 $\zeta_{\alpha\gamma} \equiv \sum_{\beta} \exp \left(-{(E_{\alpha\beta} -
E_{\gamma\beta})^2\over 4\lambda\hbar^2}\right)$ $(tr_E
\Pi_{\beta}\rho_E(0)) < 1$. The sum over $\beta$ in equation (4.7)
is an almost periodic function [25,26] that appears also for the
so-called 'correlation amplitude' in quantum decoherence theory
[27].

Now, in analogy with the proof of lemma 3.4, for sufficiently long
time interval $[t, t+T]$, such that $t_{\circ} \in [t, t+T]$, for
$\alpha\neq\gamma$, for large number of summands (many-particle
environment $E$) in equation (4.7): (a) the time average on the
interval $\lim_{T\to\infty} \langle B\rangle_T = 0$, and (b) the
standard deviation on the interval $\lim_{T\to\infty} \langle
\vert B \vert^2\rangle_T = 0$ for typical models [13] of the
many-particle $E$ system. Since $\zeta <  1$, equation (4.5) is
proved, with simultaneous observation that $t_{\circ}$ is of the
order of 'decoherence time' denoted $\tau_{dec}$  [13,15]. \hfill
Q.E.D.

Lemma 4.1 of this section is formally equivalent with Lemma 4.1(i)
of Ref. [1]: For most of the large values of $t_{\circ}$, the
r.h.s. of equation (4.7) is negligible (for $\alpha\neq\gamma$)
already for the time intervals of the order of 'decoherence time'
with the unique steady state on the r.h.s. of equation (4.5) that
is trivially divisible, while for arbitrarily large $t_{\circ}$
there is unavoidable recurrence of the initial values [18,27]. The
larger the environment $E$ the shorter 'decoherence time' and the
longer the recurrence time-interval.

As distinct from Lemma 4.1(i) of Ref. [1], equations (4.4) and
(4.5) {\it reveal how works the subsystem's dynamical map}:
divisibility (and therefore Markovianity) does not apply for
arbitrary time instants, but [in contrast to equation (3.15)] for
most of the sufficiently large values of $t_{\circ}$--again
exhibiting a need for the coarse graining of 'time axis'.

\bigskip

{\bf 4.2 The approximate map}

\bigskip

From both equation (3.21) and equation (4.2), while bearing in
mind equation (4.1) i.e. the exchange $P_m \to
P_{\alpha}\Pi_{\beta}$, the subsystem's approximate map:

\begin{equation}
\rho_S(t_{\circ}) \approx \sum_{\alpha} P_{\alpha}
\rho_S(0)P_{\alpha} + \sum_{\alpha} \left(\sum_{\beta}
\exp\left(-{\imath t_{\circ}\delta_{\alpha\beta}\over\hbar}\right)
tr_E \Pi_{\beta}\rho_E(0)\right) P_{\alpha} \rho_S(0)
\Pi^{(\alpha)} +h.c.
\end{equation}

\noindent where
$\Pi^{(\alpha)}=\sum_{\nu_{\alpha}}P_{\nu_{\alpha}}$,
$\delta_{\alpha\beta} \approx E_{\alpha\beta} -
E_{\nu_{\alpha}\beta}$  and in analogy with equation (3.20):

\begin{equation}
\exp \left(-{(E_{\alpha\beta} - E_{\gamma\beta})^2\over
4\lambda\hbar^2}\right) \ll 1,
\end{equation}

\noindent for certain set of the $\gamma$ indices for every fixed
$\alpha$. As well as equation (3.20), equation  (4.9) is typically
not applicable to the few-particle systems.

The subsystem's projectors $P_{\alpha}$ and $\Pi^{(\alpha)}$ in
equation (4.8) can be easily shown to satisfy the algebra
established for the $P_m$ and $\Pi^{(m)}$ projectors, which appear
in equation (3.21). The map is obviously linear, positive, trace
preserving and unital while satisfying equation (3.7).

The sums over $\beta$ in equation (4.8) are  almost-periodic
functions that are essentially discussed in the proof of lemma
4.1. Hence for the time intervals of the order of decoherence
time, the approximate map equation (4.8) returns the (approximate)
steady state established by lemma 4.1. Thereby  equation (4.9)
regards the coarse-grained eigenvalues (for certain choices of
$\alpha$ and $\gamma$) of the 'measured' subsystem's observable
whose eigenprojectors are exactly the $P_{\alpha}$s appearing in
equation (4.1) [28]. In the context of the decoherence theory,
such observable is often referred to as 'pointer observable' [27].

Due to the fact that $\rho_S(t_{\circ}) \approx \sum_{\alpha}
P_{\alpha} \rho_S(0)P_{\alpha}$ for most of large values of
$t_{\circ}$, one may tempt to think that the map acquires
divisibility and also complete positivity in the same time
interval. However, this is  naive as we are going to
demonstrate--the approximations useful for the quantum states need
not be as useful for the system's dynamical map. In the rest of
this section we more precisely determine the lower time bound for
'decoherence time' for equation (4.8).

Bearing in  mind that the algebra of the $P_{\alpha}$ and
$\Pi^{(\alpha)}$ projectors is formally the same as for the $P_m$
and $\Pi^{(m)}$ projectors of Section 3(b), it is now
straightforward to prove, in analogy with lemma 3.5, that the
subsystem's approximate map equation (4.8) is divisible for every
$t_{\circ}$.

In regard of complete positivity, it is  straightforward to repeat
the proof of lemma 3.4 to obtain in analogy with equation (3.25)
[we also use notational analogy]:

\begin{equation}
{1\over d} \sum_{\alpha}  \left( \sum_{i=1}^{g_{\alpha}} p_i
\right)^2 + {1\over d} \sum_{\alpha} \left(2\sum_{\beta}\left(tr_E
\Pi_{\beta} \rho_E(0)\right) \cos (\delta_{\alpha\beta}t_{\circ})
\right) \left(\sum_{i=1}^{g_{\alpha}} p_i\right) \left(
\sum_{\nu_i=1}^{g^{(\alpha)}} p_{\nu_i}\right)
\end{equation}

\noindent where $d$, $g_{\alpha}$ and $g^{(\alpha)}$ refer to the
open $S$ system and hence take the much smaller values for the $d$
and $g$s in equation (4.10) than in equation (3.25). Noticing that
the sum over $\beta$, which we denote
$\epsilon_{\alpha}(t_{\circ})$, in equation (4.10) is an almost
periodic function, from equation (4.10) we obtain the estimate:

\begin{equation}
{1\over d} \sum_{\alpha}  \left( \sum_{i=1}^{g_{\alpha}} p_i
\right)^2 + 2 \epsilon(t_{\circ}) g' \sum_{\alpha}
{\chi_{\alpha}\over g'd} \ge 0
\end{equation}

\noindent for sufficiently large $t_{\circ}$, in analogy with
equation (3.26);
$\epsilon(t_{\circ})=\max\{\epsilon_{\alpha}(t_{\circ})\}$ and
$\sum_{\alpha} \chi_{\alpha}/g'd\le 1$. Now all the coefficients
in equation (4.11) refer to the $S$ system, not to the closed
system of Section 3. The $g'$ in equation (4.11) can be roughly
estimated to be at least $\min\{tr \Pi_{\beta}\}$-times smaller
than the $g$ appearing in equation (3.25). Therefore we conclude
that the subsystem's approximate dynamical-map equation (4.8)
dynamically acquires complete positivity {\it much faster} than
the total (closed) system's approximate dynamical-map but with the
time bound that is determined by $g'$ and is therefore longer than
the
 decoherence time in lemma 4.1 (for which, formally, $g'=1$).

Hence the dynamical emergence of the time-coarse-grained
Markovianity of the subsystem's approximate dynamical map with the
 time bound that is determined by the  $g'$ parameter.

\bigskip

 {\bf 5. Markovianity state-domains for closed systems:
quantitative estimates}

\bigskip

From Section 4 we can learn that open systems in a
measurement-like (decoherence-like) interaction with their
environments are Markovian. Therefore, in this section, we
consider the closed many-particle systems.

Given  $\exp(-4) \approx 0.018 \ll 1$ and bearing in mind Section
2 from which $\hbar\sqrt{\lambda} > \min\{2\Delta H/\pi,$
$2(\langle H\rangle - E_g)/\pi\}$, the numerical condition

\begin{equation}
{E_m - E_n\over \min\{2\Delta H/\pi, 2(\langle H\rangle -
E_g)/\pi\}} \equiv {\pi\delta_{mn} \over 2 \min\{\Delta H,
(\langle H\rangle - E_g)\}} > 4, \forall{m\neq n\notin\{\nu_m\}}
\end{equation}

\noindent is necessary in order to satisfy equation (3.20).

From equation (5.1) we can recognize two domains of states that
allow for Markovianity and one domain not allowing for
Markovianity  as follows.

Denoting $E = E_{max} - E_g$ and setting $\Delta H = E/d$ and
$(\langle H\rangle - E_g) = E/r$, equation (5.1) implies:

\begin{equation}
\delta_{mn} \equiv {E\over k(m)} > 2.55 \min \left\{ {E\over d},
{E\over r} \right\}, \quad \forall{m}
\end{equation}

\noindent with the real $k(m),d,r>0$. Hence equation (5.2) reveals
the coarsening dependence on the initial state via:

\begin{equation}
k(m) < \max\{d,r\}/2.55, \quad \forall{m},
\end{equation}

\noindent which exhibits: Increase of $k(m)$ [less coarse-grained
spectrum] implies the decrease of $\Delta H$ or $\langle H\rangle
\to E_g$. Also the estimates follow: In order to have the
coarse-graining possible, $k(m) > 1, \forall{m}$ is necessary and
hence $\Delta H < E/2.55k(m) < E/2.56 \approx 0.39 E$ or $\langle
H\rangle = E_g + E/2.56 \approx E_g + 0.39 E$ are {\it required}.
For the initial states not fulfilling any of the constraints, the
equation  (3.21) is not valid. However, Markovianity can be
apparently restored for some large values of {\it both} $\Delta H$
and $(\langle H\rangle - E_g)$, for which the exact state equation
(3.3) may be practically indistinguishable from the state equation
(2.1), which pertains to the universal, global time. In between
these 'extremes'--small vs. large values of $\Delta H$ and
$(\langle H\rangle - E_g)$--are the states for which Markovian
dynamics equations (2.1) and (3.21) are not valid. For every
$k(m)$ the coarse graining interval $[E_m, E_m + E/k(m)]$ uniquely
determines the set of the energy eigenprojectors $P_{\nu_m}$ and
the parameters $g^{(m)}= \sum_{\nu_m} tr P_{\nu_m}$.

On the other hand,  numerical values of $k(m)$ determine the
limits of validity of the condition
$\exp(-\delta_m^2/4\lambda\hbar^2) \approx 1$, Section 3(b), that
is also necessary for equation (3.21) to be valid. Introduce the
measurement error $\delta E(m) = E/xk(m)$, where $x>1$. By
definition, $\delta_m = r\delta E_m, r\gtrsim 1$. Also by
definition, and due to operational indistinguishability of the
values from the interval $[E_m, E_m+\delta E_m]$, $\delta_m =
\left(E/k(m) -(E_m + E/xk(m))\right)/s, s\gtrsim 1$. Then easily
follows the constraint $x=(rs+1)/(1- k(m)E_m/E)$; for simplicity,
we omit dependence of $x,r,s$ on $m$. For the positive non-zero
$E_m$s, $k(m)<E/E_m$, in order for $x>0$.

To illustrate, consider a closed system of $N \gg 1$
noninteracting spin-$1/2$ particles (that can act as a
single-spin's environment [29]), with the Hamiltonian $H = \hbar
\omega_{\circ}\sum_{i=1}^N S_{iz}$ and the energy-scale
$C=\hbar\omega_{\circ}$. Energy spectrum
$\pm(N-p)\hbar\omega_{\circ}/2$ with the degeneracy ${{N}\choose
{p}}$, $p=0,1,2,...,N$, while $E = N\hbar\omega_{\circ}$. For the
initial state satisfying $\langle H\rangle = 0$, $\langle H\rangle
- E_g = -E_g = N\hbar\omega_{\circ}/2$, i.e. $r=2$ (cf. equation
(5.3)). The Markovianity constraint $\Delta H < 0.39 E$ can be
satisfied already with $d=3$, i.e. with $\Delta H =
N\hbar\omega_{\circ}/3$ since  $3/2.56\approx 1.172 > k >1$
satisfies equation (5.3). Then we can choose $k=1.71$ and
$\hbar\sqrt{\lambda}=0.7E/\pi > (2/\pi)\min\{E/2, E/3\}$ and
obtain for the Gaussian terms in equation (3.20) approximately to
equal $0.1$8. The smallest value for $x$ is for the ground energy
$E_g<0$, which gives rise to the largest measurement error $\delta
E = 1.08 E/(rs+1)$ and therefore the largest
$\exp(-\delta_m^2/4\lambda\hbar2)= 0.798$ for $r=1, s=9$. On the
other hand, for $\Delta H=0.4 E$ one obtains $d=2.5$ and hence
 $k< 2.5/2.56 < 1$, for which  the
coarse graining is not possible and therefore dynamics is
non-Markovian. However, for sufficiently large $\Delta H$ (small
$d$), we can obtain approximate validity of equation (2.1) as
follows. E.g. for the initial states $\vert \psi_{\pm}\rangle =
(\vert E_{max}\rangle \pm \vert E_g\rangle)/\sqrt{2}$, $d = r = 2$
and $k<2/2.56<1$, which says that the coarse graining is excluded.
However, equation (3.3) gives (with the choice $\lambda = 1.1
(E/\pi\hbar)^2$):

\begin{eqnarray}
&\nonumber& \sigma_{\pm}(t_{\circ}) \approx {1\over 2} [ \vert
E_{\max}\rangle\langle E_{\max}\vert + \vert E_g\rangle\langle
E_g\vert \pm 0.11 \exp(-\imath E t_{\circ}/\hbar) \vert
E_{max}\rangle\langle E_g \vert
\\&& \pm 0.11 \exp(\imath E
t_{\circ}/\hbar)\vert E_g\rangle\langle E_{\max}\vert ].
\end{eqnarray}

Then the fidelity [3] $\mathcal{F} = + \sqrt{\langle
\psi_{\pm}(t_{\circ}) \vert \sigma_{\pm}(t_{\circ})\vert
\psi_{\pm}(t_{\circ})\rangle} \gtrsim 0.745, \forall{t_{\circ}}$,
which makes the two states interchangeable (almost
indistinguishable) for certain practical purposes and thus
apparently the unitary (Markovian) dynamics of the system.

Let us now consider a system of noninteracting harmonic
oscillators (or uncoupled optical modes) with unique frequency
$\omega_{\circ}$ that is defined by the Hamiltonian $H =
\hbar\omega_{\circ}\sum_{k=1}^M (a_k^{\dag} a_k +1/2)$ and  the
characteristic energy-scale $C=\hbar\omega_{\circ}$, where appear
the creation and annihilation Bose operators. The energy spectrum
[in the physical units $\hbar\omega_{\circ}=1$] $\nu +M/2 > 0$,
where $\nu=0,1,2,3,...$ is an eigenvalue of the Hermitian
number-operator $N=\sum_{k=1}^M a_k^{\dag} a_k$; $N\vert
\nu\rangle =\nu\vert\nu\rangle$. Assuming a finite 'cutoff'
$\nu_{max}$, the finite $E = \nu_{max}\hbar\omega_{\circ}$ and
$\langle H\rangle - E_g = \langle N\rangle \hbar\omega_{\circ}$,
while $\Delta H = \Delta N \hbar\omega_{\circ}$. For comparison
with the spin-system, we consider $k=1.71$ and
$\hbar\sqrt{\lambda}=0.7E/\pi$, which gives the same estimate for
the Gaussians appearing in equation (3.20) as for the system of
spins. However, due to the non-negative energies, for the system
of oscillators, it is easy to detect the smallest $x=rs+1$ and the
largest $\delta E=0.58/(rs+1)$. Therefore the largest
$\exp(-\delta_m^2/4\lambda\hbar^2)= 0.936$ for $r=1, s=9$. For the
initial state considered for the spins system, it is easy to
obtain $r = d = 2$, equation (5.4) and the fidelity $\mathcal{F}
\gtrsim 0.745, \forall{t_{\circ}}$.

Better satisfied the Markovianity conditions, equation (3.20) and
$\exp(-\delta_m^2/4\lambda\hbar2)$ $\approx 1$, pertain to the
rather small $\Delta H$ and $\langle H\rangle - E_g$ that, in
turn, refer to the rather small energy-contents of the system. In
such cases the larger values for $k(m)$ allow the smaller
measurement-errors as well as make the equation  (3.21) better
satisfied.

Placing different 'frequencies' $\omega_{i}$ in the
self-Hamiltonians, or introducing  interactions in the system,
lead to a rather dense energy spectrum for $N \gg 1$ thus leading
to the continuous limit that is often used in the condensed-matter
physics [30]. The analysis of Markovianity is essentially the same
as presented above, depending on the (non)existence of the
negative energy values.

Therefore observation of (non)Markovian dynamics [e.g. via quantum
process tomography [3]] of closed systems depends on the initial
state, the structure of the energy spectrum and on the subtle
relations between the above introduced parameters, $k, r, s$.
Every measurement that cannot resolve the system's energies better
than the error $\delta E=\max\{\delta E_m\}$ may reveal
non-Markovian dynamics. The only option for detecting
non-Markovian dynamics of an open many-particle system is to
perform measurements in the time intervals shorter than
$t_{\circ}$, cf. Lemma 4.1 and Section 4(b).

While numerical details are model sensitive, the above
distinguished structure of the Markovianity state-domains is
universal. Thus we learn about the novel, rich behavior of closed
many-particle systems: Depending on the system's energy, an
observer can in principle detect low-energy Markovian dynamics,
the unitary-like [apparently unique time] Markovian behavior for
relatively high energies, and non-Markovian dynamics for the rest
of quantum states. In classical terms, 'high energy' can be linked
with high temperature, which is model-specific yet. Lemma 3.3
suggests that the observation requires interaction of the energy
scale much larger than $\hbar\omega_{\circ}$, for both here
regarded models.

\bigskip

 {\bf 6. Relation to the standard theory}

\bigskip

Despite the apparently strange foundational character, the LTS
dynamical map offers the following lessons regarding  ensembles of
many-particle systems.

First, in contrast to the standard wisdom [15-17], Markovian
character of the dynamics is not unconditional. On one hand it
appears in the coarse-graining description, Section 3(b). {\it Per
se} this is not entirely a new position--some kind of 'coarse
graining' is often found a basis for quasi-classical behavior of
quantum systems [1,24,31,32]. To this end, the quantitative
criteria (equations (3.20), (3.24) and (4.9)) set the new layer of
considerations that could, at least in principle, be
experimentally tested. On the other hand, complete positivity or
divisibility and hence Markovianity of the map are dynamically
established and 'should ripen'. That is, lemma 3.4 and lemma 4.1
indicate that the {\it dynamical maps themselves are dynamic}.

In the standard theory,  similar conclusions  appear while bearing
model-dependence in the context of a narrower definition of
Markovianity  in a perturbative treatment (weak interaction), see
e.g. [15,33]. Non-validity of equation (3.7) in the standard
theory implies that for   short times there always exists the
map-inverse and therefore dynamics is Markovian [16,17]. However,
results of Sections 3 and 4 are model independent,
non-perturbative and point out {\it non-Markovianity for the
arbitrarily short time interval} ($0<t<t_{\circ}$) that is unknown
to the standard theory.

Second, the exact, non-Markovian dynamics equation (4.2)
straightforwardly reproduces certain results known for a Markovian
counterpart from the standard theory. To see this, consider an
orthonormalized basis $\{\vert m\rangle\}$ that is adapted to the
orthogonal projectors $P_{\alpha}$ and choose only a subset of
states  such that $P_{\alpha} \vert m\rangle = \vert m\rangle$
implies $P_{\gamma} \vert m\rangle =0, \forall {\gamma \neq
\alpha}$. Then lemma 4.1 [that assumes strong interaction] implies
decoherence:

\begin{equation}
\lim_{t_{\circ}\to\infty} \rho_{Smn}(t_{\circ}) =
\lim_{t_{\circ}\to\infty} B_{mn}(t_{\circ}) \rho_{Smn}(0) = 0,
\forall{m\neq n},
\end{equation}

\noindent as a  non-Markovian process.

On the other hand, in the standard theory of strong interaction
('singular-coupling' limit), equation (6.1) appears for {\it
Markovian dynamics}. To see this, we refer to the Markovian master
equation (3.159) of Ref. [15] [$\hbar = 1$]:

\begin{equation}
\dot \rho_S(t) = -\imath [H_S + H_{LS}, \rho_S(t)] + \sum_{\alpha,
\beta} \gamma_{\alpha\beta} \left(A_{\beta}\rho_S(t)A_{\alpha} -
{1\over 2} \{A_{\alpha} A_{\beta}, \rho_S(t)\} \right)
\end{equation}

\noindent where the curly brackets denote the 'anticommutator'.

In order to comply with our considerations, we neglect the
commutator in equation (6.2) and due to the separable form of the
('pure decoherence') interaction equation (4.1), the commutators
$[A_{\alpha}, A_{\beta}] = 0, \forall{\alpha,\beta}$. Choose a
common eigenbasis, $\{\vert m\rangle\}$, for the set
$\{A_{\alpha}\}$. Then equation (6.2) gives rise to  the
matrix-elements, $\rho_{Smn} \equiv \langle m\vert \rho_S\vert
n\rangle$, differential equation

\begin{equation}
\dot \rho_{Smn}(t) = -{\Gamma_{mn}\over 2} \rho_{Smn}(t),
\end{equation}

\noindent where:

\begin{equation}
\Gamma_{mn} = \sum_{\alpha,\beta} \gamma_{\alpha\beta} (a_{\alpha
m} - a_{\alpha n}) (a_{\beta m} - a_{\beta n}) \ge 0,
\end{equation}

\noindent and $a_{\alpha m}$ is the $m$th eigenvalue of
$A_{\alpha}$.  The proof of non-negativity of $\Gamma_{mn}$s
directly follows from the observation that $\Gamma_{mn} = (\vec
v_{mn}, \gamma \vec v_{mn})$, where  the real vector $\vec v_{mn}
= (a_{\alpha m} - a_{\alpha n})$ while $\gamma =
(\gamma_{\alpha\beta})$ is a positive semi-definite matrix and
$\Gamma_{mm} = 0$. Hence from equation (6.3):

\begin{equation}
\rho_{Smn}(t) = \exp\left(-{\Gamma_{mn}\over
2}t\right)\rho_{Smn}(0),
\end{equation}

\noindent in quantitative agreement with equation (6.1) for
$t_{\circ} \sim 2\Gamma_{mn}^{-1}$.

Third, bearing in mind that the limit $t_{\circ} \to \infty$,
Lemma 4.1, typically regards the short time intervals of the
'decoherence time', equation (4.5) is not valid for arbitrarily
long time intervals. Rather, there is recurrence of the initial
values [18,27] and hence the repeating cycles of dynamical change
of the map, which is presented by equation (4.5). For large number
of constituent particles of the environment, this recurrence
typically occurs for very long time interval and thus the
repetition of the 'cycles' may be virtually unobservable.

Fourth, in the standard theory [15-17], existence and properties
of steady state(s) of the reduced dynamics is a complicated topic
and an active area of research. The averaged state (also known as
'ergodic average'), that is known to be a steady state for the
completely positive semigroup (Markovian) dynamical maps [15-17],
also appears to be steady for the both exact ({\it non-Markovian})
dynamics equation (3.3) and (4.2), of the  form:

\begin{equation}
\bar \sigma = \lim_{T\to\infty} {1\over T} \int_0^T
\sigma(t_{\circ}) dt_{\circ} = \sum_n P_n \rho(0) P_n.
\end{equation}

\noindent While the proof of equation (6.6) is an elementary
integration of equations (3.3) and (4.2), the steady state on the
r.h.s. of equation (6.6) is {\it unique}.

Finally, equations (3.25), (4.7) and (4.11) reveal different time
scales for the emergence of Markovianity. Equation (4.7) is
well-known from the decoherence theory and qualitatively defines
'decoherence time', $\tau_{dec}$ [18,27]. On the other hand, both
equations (3.25) and (4.11) require much longer time intervals to
reach the respective small values $g^{-1}$ [which for equation
(4.7) is of the order of 1]. In the absence of the exact
mathematical relation, we just note the chain of inequalities, $1
\ll g' \ll g$, that directly implies the chain of the respective
time scales $\tau_{dec} < \tau' < \tau$. The  inequality $g' \ll
g$, emphasized in Section 3(b), follows from

\begin{equation}
g = \max\{tr \Pi^{(m)}\} = \max\{\sum'_{\alpha,\beta} tr_S
P_{\alpha} tr_E \Pi_{\beta} \}\sim g' \max\{tr_E
\sum'_{\beta}\Pi_{\beta}\},
\end{equation}

\noindent [where the primed sums do not take all the possible
values of the summation coefficients] while bearing in mind that
$\Pi_{\beta}$s are the environmental projectors, equation (4.1).
From equation (6.7) directly follows: for a closed system of fixed
size (unique value of $\max\{tr \Pi^{(m)}\}$), smaller subsystem
is monitored by larger environment and therefore $g'$ decreases
with the decrease in the size of the  subsystem. Physically this
means that, [for the same total, closed system], the smaller the
open system the faster is reached the steady state and the longer
'recurrence time'. Observations of the open system that are not
limited by coarse-graining give rise to the fastest approach (with
the rate $\tau_{dec}^{-1}$) to the steady state.

\bigskip

 {\bf 7. Discussion}

\bigskip
Dynamical map imposed by Local Time Scheme is mathematically of a
rather special kind that is here introduced for the first time. In
certain points, Local Time Scheme fits with, goes beyond, or
extends the results of the standard open systems theory. Thereby
Section 6 answers the first question on usefulness of Local Time
Scheme that is posed in Introduction. Within the scope of the new
fundamental dynamical law equation (2.2)--that takes the place and
the role of the Schr\" odinger law, equation (2.1), of the
standard theory--, the findings of Sections 3 and 4 are {\it
generic} and {\it universal}. Here we specifically emphasize
non-existence of the short-time Markovianty, the novel, rich
behavior of closed many-particle systems and that the scheme
provides a long-sought answer to the problem of microscopic origin
of the phenomenological rule, that the smaller open systems relax
[in our case: attain the steady state] faster than the larger
ones.

We conclude that dynamics imposed by Local Time Scheme
successfully though specifically tackles the so-far-investigated
topics in the foundations of quantum theory. Nevertheless, there
remains yet much to be done. We can detect at least the following
four research directions. First, experimental tests of our
findings are methodologically on the same ground as the tests of
the occurrence of decoherence. To this end, a rather precise
knowledge of the energy spectrum as well as sophisticated control
of quantum states of many-particle systems are needed. Beyond
these technical requirements, we do not see any obstacles to the
more-or-less straightforward experimental tests of our results.
Second, usefulness of Local Time Scheme regarding the
interpretational issues of quantum theory (e.g. of quantum
measurement) should be separately investigated. In this regard, of
particular interest is description of a {\it single} system
dynamics  [as distinct from the ensemble of systems considered so
far]. Third, implications for the physical nature of time allowed
by LTS may contribute to the debate of whether Time is fundamental
or not and so whether or not the task of spacetime quantization
makes any sense. Finally, in certain points of the foundational
character, Local Time Scheme concurs with a cosmological program
that has recently been reviewed by Hartle [34]. Here we briefly
emphasize certain essential points of the program that have clear
counterparts within LTS while leaving details for the sequel.

The low entropy initial state of the Universe [34] is directly
recognized within LTS: Due to the points (a)-(e) in Appendix A,
the early Universe is a single 'local' system that is subjected to
unique local time and therefore described by equation (2.1) for
pure
 i.e. for the zero-entropy states.
Subsequent dynamics of the Universe 'brings today's nearly
isolated systems' [34], which, according to LTS, are described by
their own local times and mixed (non-zero entropy) states of the
form of equation (2.2) i.e. of equation (3.1). Robustness to
external noise is the characteristic trait of the
decoherence-induced 'quasiclassical' domain that requires {\it
sufficiently strong interaction} of a {\it many}-particle system
with its environment in a {\it coarse-grained} description [34].
Sections 3 and 4 provide the basis for the requirements, which are
above emphasized by italics. Finally, the conjecture '{\it A
situation of} local equilibrium {\it will generally be reached
before complete equilibrium is established, if it ever is.}" [34]
can be recognized to summarize lemma 3.2, lemma 4.1 and the two
final remarks in Section 6.

\bigskip

{\bf 8. Conclusion}

\bigskip

The so-far investigated mathematical aspects of Local Time Scheme
prove to be consistent and physically  useful. The dynamical map
imposed by the Scheme is mathematically of a special kind that
physically quantitatively fits with, goes beyond, or extends the
results of the standard open quantum systems theory and
qualitatively tackles certain foundational issues in cosmology.
The interpretational implications of the scheme, such as quantum
measurement problem and the transition from quantum to classical,
are yet to be investigated.

\bigskip

{\bf Appendix A}

\bigskip

In Local Time Scheme, the primitive is {\it dynamics}, which is
simply  a map between the system's states:

$$\rho_1 \to \rho_2 \to \rho_3 \to... \eqno(A.1)$$

Along the dynamical chain equation (A.1), the system can be
subjected to
 changes of its local time, which can be shared by some other
systems that, as a whole, constitute an at least
approximately-closed system subjected to the unitary Schr\"
odinger dynamics. For a single closed system, the states are pure
($\rho_i^2 = \rho_i$) and the chain equation (A.1) is a 'history',
which assumes unique local time [35]. Therefore, in a composite
system, such as the Universe itself, there may be more than one
such closed ('local') system described by its own local time.
Hence  a composite system is described solely by the composite
system's Hamiltonian, which defines [dynamically changing]
distribution of local systems and their local times without
'history' in the standard sense [35].

The rules for determining local time  are as follows [1]: (a)
Systems with different Hamiltonians such as those with a different
numbers of particles, or different kinds of particles, or
different kinds of interactions between the particles are subject
to different local times; (b) Systems that mutually interact are
subjected to the same time; (c) Noninteracting systems need not
have a common time; (d) Nonidentical many-body systems which do
not interact and locally follow independent Schr\" odinger
dynamics do not have a common time--which makes the universal time
undefinable; (e) Local time refer even to the mutually identical
many-body systems, as long as they represent the mutually
independent local systems.

These rules constitute the foundational basis for Local Time
Scheme. Everything one might need is 'inscribed' in the
Hamiltonian since, along the dynamical chain equation (A.1), the
average value  of every term of the  Hamiltonian and the
energy-scale $C$ are uniquely determined. For a closed system,
unique Hamiltonian determines unique energy scale $C$. On the
other hand, interaction may change the energy scale for the total
system's Hamiltonian. If interaction for a pair of closed systems
is weak, the systems are approximately independent and
approximately subjected to independent local times. This should be
sharply distinguished from the case of 'weakly interacting
systems' that are subsystems of a local (approximately closed)
system; such cases  are considered in the context of Markovian
dynamics in the standard open systems theory [15-18]. The other
'extreme' is the sufficiently strong interaction which defines
formation of a new composite system and the start ($t=0$) of the
new local time ticking; such situations are considered in Lemma
3.3 and Section 4 of the body text. In between the two 'extremes'
are the cases for which the interaction does not dominate neither
is sufficiently weak.

For a sub-chain $\rho_i\to \rho_{i+1}\to \rho_{i+2}\to...$ of the
dynamical chain equation (A.1), we say to be of {\it fixed energy
scale} if the respective energy scales, $C_i, C_{i+1},
C_{i+2}...$, are approximately equal, i.e. if $C_i \approx C_{i+1}
\approx C_{i+2}... = C$.

Local Time Scheme does not provide quantitative {\it kinematical}
criteria for 'weak' vs. 'sufficiently strong' interaction.
Therefore the following question is in order:
 Are there any realistic (at least approximately) closed
many-particle systems and how can they be recognized in the
realistic physical situations? This concern raised in Section 2
is, in our opinion, purely operational.

Nonexistence of at least approximately closed systems as described
above would be fatal for most of the physical theories as well as
for interpreting certain experimental tests. For the whole of the
decoherence and measurement theory, as well as for the foundations
of the open systems theory, validity of the Schr\" odinger law
equation (2.1) is essential and is implemented by 'object of
measurement + apparatus (+ the apparatus' environment)' as well as
by '[open]system + environment' [15-18]. Approximate isolation
from the rest of the Universe is indispensable also for the
appearance of the energy band-structure of electrons in crystals
[30], for the so-called 'alpha-clustering' in nuclear physics [36]
(and references therein), 'entanglement renormalization' [37], as
well as for existence of quasiclassical realms in cosmology [34]
etc. Therefore we do not regard the above-posed question
foundational but rather of the more-or-less practical, technical
nature that should be separately answered for every concrete
physical situation.

The second concern raised in Section 2 refers to the sharp {\it
dynamical} transition from the  'weak' to 'strong' interaction of
the initially independent local systems. For the unsharp
transition, the initial instant $t=0$ of the common local time for
interacting systems {\it a priori}, i.e. independently of Lemma
3.3 [which concerns the fixed energy scales], is not well defined.
Nevertheless, this does not produce any problem as long as
dynamics equation (A.1) can be 'coarse grained' in order to
provide a 'quick' dynamical transition from the weak to the
sufficiently strong interaction.

Consider a dynamical chain

$$\rho_i\to\rho_{i+1}\to ...\rho_{i+k}\to...
\rho_p\to\rho_{p+1}\to ...\rho_{p+q} \eqno(A.2)$$

\noindent where the sub-chains $\rho_i\to\rho_{i+1}\to
...\rho_{i+k}$ and $\rho_p\to\rho_{p+1}\to ...\rho_{p+q}$ are with
the fixed energy scales, $c$ for the self-Hamiltonian and $C$ for
the interaction term, respectively. Then reducing the sub-chain
$\rho_{i+k}\to... \rho_p$ to arbitrary {\it single state} of the
sub-chain is the dynamical coarse graining. By definition, this
coarse graining assumes {\it practical indistinguishability} of
certain dynamical\-ly-close states and is finer than and embedded
in the dynamical coarse graining necessary for
 the dynamical appearance of
Markovianity, Sections 3 and 4.

Needless to say, the coarse graining is equally relevant for the
strongly-interacting systems in the standard theory of global
time: Indeed, it is not {\it a priori} clear how quickly the
interaction becomes sufficiently strong and whether or not the
initial ensemble is really pure as assumed by equation (2.1). The
plausible answer provided below Lemma 3.3 in the body text, that
$U(\delta t) \approx I$, is encompassed by the state
 coarse graining; going beyond this approximation, one can find e.g. [38].
 Therefore the dynamical coarse graining generally appears
 to be as necessary as it is of the operational nature and thus
not producing any problems in the foundations and application of
Local Time Scheme.

[1] Jekni\' c-Dugi\' c J, Arsenijevi\' c M \& Dugi\' c M. 2014.  A
local-time-induced pointer basis. {\it Proc. R. Soc. A} {\bf 470},
20140283.

[2] Kitada H. 1994. Theory of local times, {\it Il Nuovo Cim.}
{\bf 109 B}, 281.

[3] Nielsen MA \& Chuang IL. 2000 {\it Quantum Computation and
Quantum Information}. Cambridge: Cambridge University Press.

[4] Milburn GJ. 1997.  {\it Schrodinger's Machines: The Quantum
Technology Reshaping Everyday Life}. W. H. Freeman.

[5] Ramsden J. 2011. {\it Nanotechnology: An Introduction (Micro
and Nano Technologies)}. William Andrew.

[6]  Durr D, Goldstein S \& Zanghi N. 2013.  {\it Quantum physics
without quantum philosophy}. Berlin: Springer.

[7] Wallace D. 2012. {\it Emergent Multiverse: Quantum Theory
According to the Everett Interpretation}. Oxford: Oxford
University Press.

[8] Kastner RE. 2012. {\it The Transactional Interpretation of
Quantum Mechanics: The Reality of Possibility}. New York:
Cambridge University Press.

[9] Fuchs CA, David Mermin N \& Schack R. 2014.  An Introduction
to QBism with an Application to the Locality of Quantum Mechanics,
{\it Am. J. Phys.} {\bf 82}, 749.

[10] Saunders S, Barrett J, Kent A \& Wallace D.  (Eds.). 2010.
{\it Many Worlds? Everett, Quantum Theory, and Reality}. Oxford:
Oxford University Press.

[11] Brunner N, G\" uhne O \& Huber M (Eds.). 2014. {\it A Special
issue on 50 years of Bell's theorem}, J. Phys. A: Math. Theor.
{\bf 47}.

[12] First iWorkshop on the Meaning of the Wave Function,
http://www.ijqf.org/groups-2/meaning-of-the-wave-function/forum/

[13] John Bell Workshop 2014,
http://www.ijqf.org/groups-2/bells-theorem/forum/

[14] Quantum Foundations Workshop 2015,
http://www.ijqf.org/groups-2/quantum-foundations-workshop-2015/forum/

[15] Breuer H-P \& Petruccione F. 2002. {\it The theory of open
quantum systems}. Oxford University Press.

[16] Rivas \' A \& Huelga SF. 2011. {\it Open quantum systems: an
introduction}. Berlin: Springer.

[17] Rivas \' A, Huelga SF \& Plenio MB. Quantum non-Markovianity:
characterization, quantification and detection, {\it Rep. Prog.
Phys.} {\bf 77}, 094001.

[18] Joos E, Zeh H-D, Kiefer C, Giulini DJW, Kupsch J \&
Stamatescu I-O. 2003. {\it Decoherence and the appearance of a
classical world in quantum theory}. (second edition) Berlin:
Springer.

[19] Petrat S \& Tumulka R. 2014. Multi-time Schr\" odinger
equations cannot contain interaction potentials. {\it J. Math.
Phys.} {\bf 55}, 032302.

[20] Margolus N \& Levitin LB. 1999. The maximum speed of
dynamical evolution. {\it Physica D} {\bf 120}, 188.

[21] Dugi\' c M \& \' Cirkovi\' c MM. 2002. Quantum information
processing: the case of vanishing interaction energy, {\it Phys.
Lett. A} {\bf 302}, 291.

[22] Audenaert KMR \& Scheel S. 2008. On random unitary channels,
{\it New J. Phys.} {\bf 10}, 023011.

[23] Jamiolkowski A. 1972. Linear transformations which preserve
trace and positive semidefiniteness of operators. {\it Rep. Math.
Phys.} {\bf 3}, 275.

[24] Jeong H, Lim Y \&  Kim MS. 2014. Coarsening measurement
references and the quantum-to classical transition. {\it Phys.
Rev. Lett.} {\bf 112}, 010402.

[25] Kac M. 1943. On the distribution of values of trigonometric
sums with linearly independent frequencies. {\it Am. J. Math.}
{\bf 65}, 609.

[26] Besicovitch AS. 1963. {\it Almost periodic functions}. Dover:
Cambridge University Press.

[27] Zurek WH. 1982. Environment-induced superselection rules.
{\it Phys. Rev. D} {\bf 26}, 1862.

[28] Dugi\' c M. 1997. On diagonalization of a composite-system
observable. Separability. {\it Phys. Scr.} {\bf 56}, 560.

[29] Zurek WH. 1981. Pointer basis of quantum apparatus: Into what
mixture does the wave packet collapse? {\it Phys. Rev. D} {\bf
24}, 1516.

[30] Kittel C. 2004. {\it Introduction to Solid State Physics}.
{eighth edition) New York: Wiley.

[31] Von Neumann J. 1955. {\it Mathematical foundations of quantum
mechanics}. Princeton: Princeton University Press.

[32] Omn\' es R. 1994. {\it The interpretation of quantum
mechanics}. Princeton: Princeton University Press.

[33] Breuer HP, Kappler B \& Petruccione F. 2001. The
Time-Convoluti\-on\-less Projection Operator Technique in the
Quantum Theory of Dissipation and Decoherence, {\it Ann. Phys.
(N.Y.)} {\bf 291}, 36.

[34] Hartle J. 2010. Quasiclassical realms, in: S.  Sounders et al
(Eds.) {\it Many Worlds? Everett, Quantum Theory, and Reality}.
Oxford: Oxford University Press, p 73.

[35] Griffiths RB. 2002. {\it Consistent Quantum Theory}.
Cambridge: Cambridge University Press.

[36] Kanada-En'yo Y. 2015. Analysis of delocalization of clusters
in linear-chain alpha-cluster states with entanglement entropy.
{\it Phys. Rev. C} {\bf 91}, 034303.

[37] Evenbly G \& Vidal G. 2014. Real-Space Decoupling
Transformation for Quantum Many-Body Systems. {\it Phys. Rev.
Lett.} {\bf 112}, 220502.

[38] Bonifacio R, Olivares S, Tombesi P \& Vitali D. 2000. A model
independent approach to non dissipative decoherence. {\it Phys.
Rev. A} {\bf 61}, 053802.

\end{document}